\documentclass[]{aastex631}

\usepackage{rotating}
\usepackage{graphicx}
\usepackage{overpic}

\DeclareUnicodeCharacter{2212}{-}

\shorttitle{Source Surface Height Optimisation for Improved Solar Wind Velocity Forecasting. }
\shortauthors{Kumar et al.}
\graphicspath{{./}{figures/}}

\begin{document}
\title{Source Surface Height Optimisation for Improved Solar Wind Velocity Forecasting Across Solar Cycles 23, 24 and 25.} 

\author[0000-0002-3902-5526]{Sandeep Kumar}
\affiliation{Udaipur Solar Observatory, Physical Research Laboratory, Udaipur, 313001, India}
\affiliation{Discipline of Physics, Indian Institute of Technology Gandhinagar, Palaj, Gandhinagar-382 355, Gujarat, India}

\author[0000-0002-0452-5838]{Nandita Srivastava}
\affiliation{Udaipur Solar Observatory, Physical Research Laboratory, Udaipur, 313001, India}

\author[0000-0002-9311-9021]{Dana-Camelia Talpeanu}
\affiliation{ Solar-Terrestrial Centre of Excellence—SIDC, Royal Observatory of Belgium, 1180 Brussels, Belgium}

\begin{abstract}

The potential field source surface (PFSS) model is a cornerstone of many state-of-the-art space weather forecasting frameworks. It includes a single physical free parameter, the source surface (SS) height. Traditionally, a fixed SS height of 2.5 $R_\odot$ has been adopted in most studies.
However, several studies have investigated the SS height in the PFSS model using various datasets and temporal intervals. Earlier efforts primarily aimed to improve estimates of the heliospheric open magnetic flux. Our recent study \cite{kumar:2025}, showed that optimising the SS height substantially improves the performance of the Wang–Sheeley–Arge (WSA) solar wind speed prediction model at L1. That analysis employed two types of GONG synoptic magnetic field maps, namely standard (STD) and zero-point-corrected (ZPC) maps, over selected periods of solar cycles (SCs) 24 and 25. In this study, we perform a comprehensive investigation of SCs 23 to 25 using synoptic magnetic field maps derived from ground-based (GONG) and space-based (SDO/HMI) observatories. Our results demonstrate that optimising the SS height in the PFSS model significantly improves the WSA model performance compared to the commonly adopted fixed SS height. We find that during solar minimum, higher SS heights ($\geq 2.5~R_\odot$) yield better solar wind speed predictions at L1, whereas during active phases the opposite trend is observed. Although the absolute optimised SS height varies across SCs, the overall distribution pattern remains similar. These finding suggests a relationship between SS height and SC phase over long time scales. Overall, the SS height optimisation enhances the background solar wind speed prediction models in the heliosphere.

\end{abstract}

\section{Introduction} 
\label{sec:intro}

Current state-of-the-art physics-based operational space weather forecasting frameworks rely on magnetic field extrapolation models \citep{arge:2000,enlil:2004,pomel:2018,kumar:2020,Mayank_2022}. These frameworks couple coronal magnetic field models with heliospheric models to predict solar wind properties at L1. In general, the Potential Field Source Surface \citep[PFSS;][]{sch:1969} extrapolation model, which provides the large-scale coronal magnetic field configuration, is used as an input to empirical solar wind models. The PFSS model represents a first-order, minimum-energy approximation of the Sun’s magnetic field.\\
The source surface (SS) height is the only physical free parameter in the PFSS model and is crucial for the extrapolation of the field above the photosphere.  
Using the PFSS model,  \cite{hoek:1983} reconstructed the heliospheric magnetic field structure during the rising, maximum, and early declining phases of solar cycle (SC) 21 (1978-1982). By comparing the in-situ IMF polarity with the modeled IMF using the PFSS model, they suggested an optimal SS radius close to $2.5\,R_\odot$. Thereafter, $2.5\,R_\odot$ has been commonly used as a conventional SS height in the community \citep{Riley:2015, reiss:2019, kumar:2020, Narchen:2020, kumar:2022, Mayank_2022}. The SS height significantly affects both the position of modeled coronal hole boundaries and their overall morphology \citep{link:1999}. Originally, SS height was defined as the coronal height at which the thermal energy density in the coronal environment exceeds the transverse magnetic energy density \citep{sch:1969}. As the Sun goes through different phases during an SC, therefore the SS height in the PFSS model should vary with the phase of the SC, which in turn controls the various modeled outputs of the model.\\

Table~\ref{tab:studies} lists various studies about SS height optimisation regarding different modeled parameters using PFSS model. It is worth noting that previous studies can broadly be classified based on the following two categories. The first distinction can be based on the temporal coverage of the analysis, i.e., whether the study is short-term, spanning only a few months, or long-term, extending over several years. The second distinction can be based on the PFSS model output, which is compared with the heliospheric observations, including unsigned and signed open magnetic flux, coronal footpoints on the Sun, magnetic field line orientations, streamer locations, and the predicted solar wind velocity at L1. Since the PFSS model represents an average solar magnetic field, and because different physical quantities are sensitive to different choices of the SS height, variations in the SS height can lead to significantly different results \citep{lee:2011}.\\

Using synoptic maps of SC22 and SC23 from Mount Wilson Observatory (MWO) and Solar and Heliospheric Observatory's Michelson Doppler Imager \citep[SOHO/MDI;][]{mdi:1995}, \cite{lee:2011} showed that the conventional SS height of 2.5 $R_\odot$ is not always optimal for the PFSS model, based on its prediction of open magnetic flux and coronal holes on the surface of the Sun. Using similar proxies, \cite{Arden:2014} tested different SS height values through SC23–24 and found that raising the SS height by 15–30$\%$ during solar minimum can lead to a better agreement of the modeled open flux with the observations. Later, \cite{nik:2019} examined PFSS solutions using GONG standard maps and explicitly showed how open flux and coronal hole areas depend on the SS height. They also reported that using a lower value of SS height during the SC24 maximum improves the PFSS model forecasting accuracy of the open flux at L1.

Apart from the above-mentioned studies, \cite{zhang:2023} optimized the SS height in the context of the WSA solar wind model for only two selected Carrington rotations (CRs) and reported that adopting lower and higher SS heights than 2.5 $R_\odot$ during solar maximum and minimum, respectively, significantly improves the WSA-derived solar wind speed at L1. More recently, our extended detailed analysis based on 16 selected CRs in SC24 and SC25 \citep{kumar:2025} using three types of GONG synoptic magnetic field maps demonstrated that using a lower and higher value of SS height compared to conventional SS height, during SC maximum and minium respectively, improves the solar wind velocity prediction at L1. We also reported that GONG ZPC maps out perform the GONG STD maps. Although most of the above-mentioned studies \citep{lee:2011,Arden:2014,zhang:2023,kumar:2025} differ in their estimates of the optimal absolute SS height, they consistently agree on the relative variation of the best-fit SS height with the phase of the SC. 

  \cite{Bena:2024}  and  \cite{wag:2022} compared large-scale coronal features observed during solar eclipses with PFSS extrapolations. Using 10 solar eclipse observations spanning 2008-2020, \cite{Bena:2024} reported optimal SS heights of 1.3 $R_\odot$ and 3.0 $R_\odot$ during solar maximum and minimum conditions, respectively. On the other hand, \cite{wag:2022} reported a nearly constant optimal SS height of 2.4 $R_\odot$ based on two solar eclipse observations, on August 1 2010 and July 11 2012. Furthermore, the PFSS model is often coupled with the Schatten Current Sheet \citep[SCS;][]{sch:1971} model. Additionally, \cite{asv:2019} investigated the SS height when the PFSS model was coupled with the SCS model in the EUropean Heliospheric FORecasting Information Asset \citep[EUHFORIA;][]{pomel:2018} framework. Their results showed no clear relationship between the SS height and the phase of SC.

\cite{huang:2024} compared PFSS maps with the Alfv\'en Wave Solar Model \citep[AWSoM;][]{awsome:2014}  MHD model to determine the SS height that best reproduces open-field regions. They concluded that the optimal SS height may be lower than 2.5 $R_\odot$ during solar minima and slightly higher than 2.5 $R_\odot$ near solar maxima. Recently, \citet{shoda:2025} empirically optimised the SS height in PFSS extrapolations by comparing modeled open magnetic flux with in-situ measurements at 1~AU over nearly two SCs. Using synoptic magnetic field maps from multiple observatories, like Solar Dynamics Observatory's Helioseismic and Magnetic Imager \citep[SDO/HMI;][]{hmi:2012}, they demonstrated that the optimal SS height varies with solar activity and the large-scale structure of the photospheric magnetic field, showing stronger correlations with the mean unsigned field strength and magnetic dipolarity than with the sunspot number. These results are in contrast to the findings reported by earlier studies discussed above, i.e., \cite{lee:2011,Arden:2014,zhang:2023,kumar:2025} .\\

Only a limited number of earlier studies examined SS height optimisation within the PFSS model in the context of its use in the WSA. Moreover, these investigations are confined to short-term analyses, at different phases of the SC \citep{zhang:2023}. Although we recently conducted a detailed analysis of 16 CRs using different types of GONG synoptic magnetic field maps spanning various phases of SCs 24 and 25 \citep{kumar:2025}, a comprehensive long-term investigation of SS height optimisation within the WSA model is still lacking. In order to resolve this issue and to have an improved understanding, the present comprehensive work incorporates synoptic magnetic field maps from space-based (SDO/HMI) and ground  observatory (GONG). It covers the GONG data from 2006 to 2025, i.e., some part of SC23, the complete duration of SC24 and SC25, and the HMI synoptic magnetic field maps from 2010 to 2025, covering SC24 and SC25.
Our primary objective is to investigate the variation of the SS height within the PFSS framework, particularly in view of its crucial role in the Wang–Sheeley–Arge (WSA) model \citep{Riley:2001,wsa_2003,Riley:2015}. 
\begin{table*}
\centering
\small \caption{Overview of studies focused on optimising the SS height. LT and ST represent long-term and short-term analyses, respectively. The proxy corresponds to PFSS-modeled parameters, which were compared with observational data.}
\begin{tabular}{|c|c|c|c|c|}
\hline
\textbf{Study} & \textbf{Type} & \textbf{Input Magnetogram} & \textbf{Proxy} &  \textbf{$R_{ss}$($R_\odot$)trend in SC Min/Max}\\
\hline
 \cite{hoek:1983}& LT & WSO & Polarity of IMF (at 1 au) &
Fixed 2.5 \\ \hline
\cite{lee:2011}& LT & MWO and MDI & open flux & higher/lower\\ \hline 
\cite{Arden:2014} & LT & LMSAL (MDI) & open flux &higher/lower \\ \hline 
\cite{nik:2019} & LT & GONG & open flux  & higher/lower\\ \hline 
\cite{asv:2019}  & ST & ADAPT-GONG & CH Characteristics & no trend for 15 CHs\\ \hline 
\cite{bad:2020} & ST & ADAPT-GONG $\&$ HMI & polarity  observed at SolO & 1.5 \\ \hline
\cite{wag:2022}  & ST & GONG and HMI & coronal features & 2.4 (10 case studies) \\ \hline 
\cite{zhang:2023} & ST  & GONG ZPC & SW velocity at L1 & higher/lower \\ \hline
\cite{huang:2024} & ST & ADAPT-GONG & open-field area & lower/higher  \\ \hline
\cite{Bena:2024} &LT& ADAPT & total solar eclipse Data &  higher/lower (10 case studies)   \\ \hline
\cite{kumar:2025} & ST & GONG ZPC and STD & SW velocity at L1 & higher/lower  \\ \hline
\cite{shoda:2025} & LT &HMI, MDI and GONG-ADAPT & unsigned open flux & inconclusive   \\ \hline
\end{tabular}
\label{tab:studies}
\end{table*}

\section{Data And Methodology}
\label{sec:data}
 We selected photospheric synoptic magnetic field maps from the Global Oscillation Network Group (GONG) and the HMI instrument. It is important to note that GONG STD and ZPC maps \footnote{\url{https://magmap.nso.edu/archive.html}} are available since August 2006, while HMI radial synoptic maps\footnote{\url{http://jsoc.stanford.edu/HMI/LOS_Synoptic_charts.html}} have been accessible since 2010. The resolution of the HMI synoptic magnetic field maps is $3600\times 1440$, whereas GONG synoptic magnetic field maps  have a resolution of $360\times180$ (on cylindrical equal area projection). To compare the results obtained from HMI synoptic magnetic field maps with those from GONG synoptic magnetic field maps, in this work, for simplicity, we have reduced the resolution of the HMI synoptic magnetic field maps to match that of the GONG synoptic magnetic field maps. GONG provide synoptic magnetic field maps in two varieties, the first is the hourly updated maps, and the second are integral full Carrington rotation maps. In our previous work, we found that hourly-updated synoptic magnetic field maps from GONG yield results similar to those from integrated full Carrington synoptic magnetic field maps \citep{kumar:2025}.  Therefore, in the present work, we used full CR synoptic magnetic field maps from GONG and HMI.

Figure~\ref{fig:ssn} presents the monthly averaged sunspot number spanning SC23, SC24, and SC25. Monthly sunspot number data were obtained from the SILSO database\footnote{https://www.sidc.be/SILSO/} to identify different phases of the SC and to select the corresponding study intervals.
The blue vertical line indicates the time of the earliest availability of STD and ZPC synoptic maps from the GONG archive, while the orange dashed vertical line marks the availability of the first HMI radial synoptic maps.
In the present work, we divide the overall time period of our study into distinct intervals as mentioned in Table \ref{tab:time} to analyze the results based on the level of solar activity and their relevance to our previous study \citep{kumar:2022}.
The light grey forward slashed region denotes the deep minimum phase of SC23. During this period, we analyzed the exceptionally low area of low- to mid-latitude coronal hole \citep{kumar:2022} area using SOHO/EIT synoptic coronal hole maps \cite{Hamada:2019}.
The lavender shaded region represents the period of maximum activity during SC24 and SC25, whereas the  backward-slashed region within SC24 highlights the interval of significantly reduced activity during the SC24 maximum. These two intervals have important implications for the overall results, as discussed in section \ref{sec:data}. \\
To model the solar wind velocity profile in the heliopsphere, in this work we adopted the same methodology as that described in our previous work  \citep{kumar:2025}. For each Carrington rotation, we start with the photospheric synoptic magnetic field maps,  and then we extrapolate the coronal magnetic field, using the PFSS model up to the SS height in the corona. In this work, we consider eight different values of the SS height, ranging from 1.5 $R_\odot$ to 3.25 $R_\odot$ in steps of 0.25 $R_\odot$. Compared to our previous study, where only three discrete SS heights (2.0 $R_\odot$, 2.5 $R_\odot$, and 3.0 $R_\odot$) were examined \citep{kumar:2025}, an extended range and finer sampling adopted in the present work enable a detailed characterization of the distribution of optimized SS heights.
We would like to mention that the range of the SS height around the conventional SS height in our study is somewhat arbitrary. However, it is consistent with the range of SS heights considered by \cite{asv:2019}, i.e., 1.4 $R_\odot$ to 3.2 $R_\odot$. 
For each value of $R_{ss}$ in the PFSS model, we extrapolate the coronal magnetic field and trace the magnetic field lines between the photosphere and the SS.  We calculate the properties of the open sub-Earth field lines, i.e., the expansion factor ($f_s$) and the minimum angular distance in radians from the nearest coronal hole boundary ($\theta_b$).  We then estimate the empirical solar wind velocity at the SS using the WSA model given by \cite{Riley:2015}.
\begin{equation}
    v_{sw}^{wsa} (f_s, \theta_b)= v_{slow} + \frac{v_{fast}-v_{slow}}{(1+ f_s)^{\alpha}} \left( \beta- \gamma e^{-(\theta_b /w)^{\delta}} \right)^{3.5}
   \label{ch3_eq:WSA}
\end{equation}
Here, the parameters $v_{\rm slow}$ and $v_{\rm fast}$ correspond to the velocity of the fastest and slowest solar wind stream. $\alpha$, $\beta$, $\gamma$, $\delta$ and $w$ are the tunable parameters of the model. In this work we used the following default values for the  WSA parameters: $v_{\rm slow} = 250 \text{ km/s}$ and $v_{\rm fast} = 750 \text{ km/s}$, $\beta = 1.0$, $\alpha = 1.5/9$, $w = 0.01$, $\gamma = 1.0$, and $\delta = 1.5$, similar to our previous work.
We further use the Heliospheric Upwind eXtrapolation  \citep[HUX;][]{[]Riley:2011} model to extrapolate the solar wind velocity profile obtained from the outer boundary of the coronal domain, i.e., from the SS, to L1. At L1, for each CR, we compared the modeled solar wind velocity profile with the observed solar wind velocity profile, for each value of the SS height selected in our study. Therefore, the combined PFSS+WSA+HUX provides a full framework to predict the solar wind velocity at L1 starting from photospheric synoptic maps.

We quantify the agreement between the observed and modeled solar wind velocity profiles at L1 using the Pearson correlation coefficient ($cc$). The coefficient $cc$ ranges between $-1$ and $1$, where values close to $1$ indicate a strong positive linear correlation, values near $0$ indicate little to no linear correlation, and values close to $-1$ indicate a strong negative correlation. We use $cc$ as the primary metric to assess model performance and discuss the results in terms of its variation across different cases. Since our objective is to determine the optimal SS height for different CRs, we define the SS height that yields the maximum value of $cc$ as the "optimized SS height" within the explored parameter space.

In-situ measurements of the solar wind speed at L1 from the OMNI database\footnote{https://omniweb.gsfc.nasa.gov/form/dx1.html} were used to validate the modeled solar wind velocity profiles at L1.
For the details on the methodology, see \cite{kumar:2025}.

\begin{table*}[ht]
\centering
\caption{Different time periods mentioned in Figure~\ref{fig:ssn}. $T_D$ is the deep minimum period of SC23 as reported by \cite{kumar:2022}.}
\begin{tabular}{c c}
\hline
Time period & Phase  \\
\hline
T1 & SC23 Descending to SC24 Ascending \\
T2 & SC24 Maximum\\
T3 & SC24 Descending+ SC24 Minimum\\
T4 & SC25 Ascending \\
T5 & SC25 Maximum \\
$T_H$ & Ascending phase of SC24 observed in HMI\\
$T_D$ & Deep minimum of SC23 inside T1\\
$T_M$& Decreased activity period during maximum of SC24\\
\hline
\end{tabular}
\label{tab:time}
\end{table*}
\section{Results and Discussion}
\label{sec:results}

\begin{figure}[htbp]
    \centering
    \includegraphics[width=\linewidth]{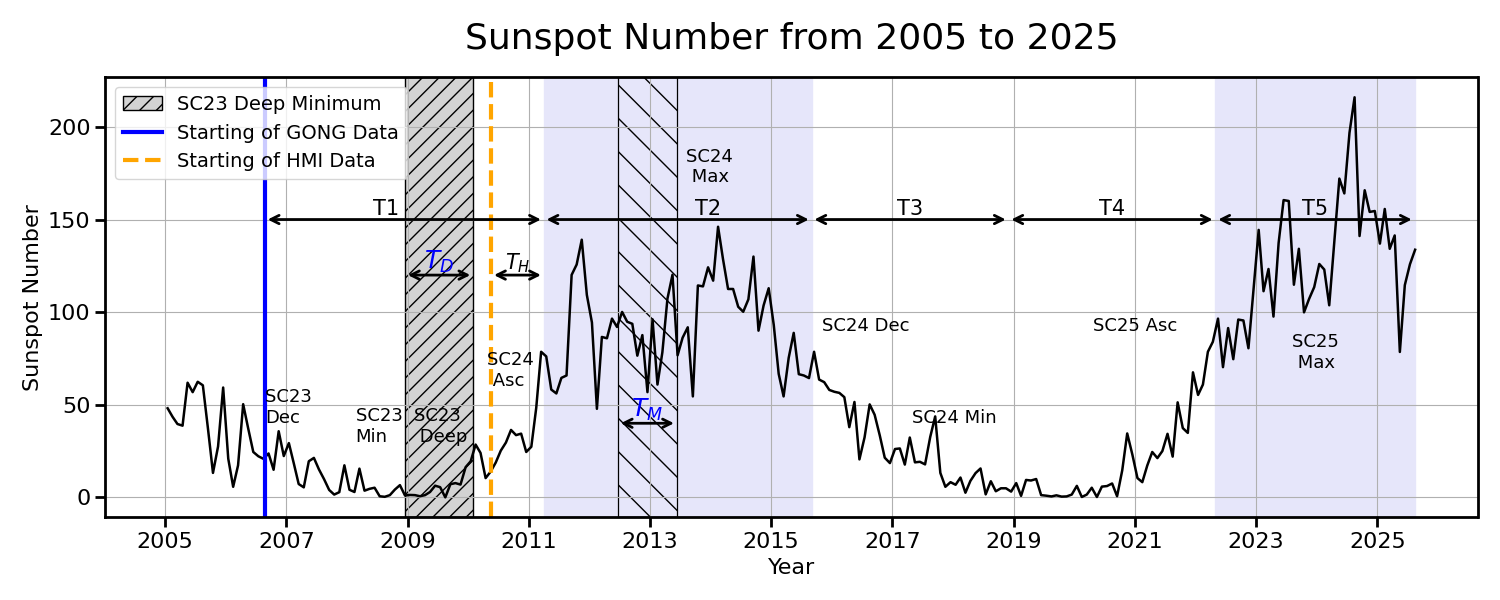}
    \caption{Monthly sunspot numbers from 2005 to 2025. The blue solid and dashed orange vertical lines mark the first available GONG and HMI synoptic magnetogram maps, respectively. Light grey forward slashed region ($T_D$) is the deep minimum period of SC23. For the details of T1, T2, T3, T4 and T5, refer to Table \ref{tab:time}. Backward slashed region ($T_M$) inside T2 marks the significantly low activity period during SC24 maximum. We used lavender colour shading for the maximum period of SC24 and SC25.}
    \label{fig:ssn}
\end{figure}
 In this section we report the results of the SS height optimisation and examine its influence on the performance of the PFSS+WSA+HUX framework across SCs 23–25. This section analyses the dependence of the optimised SS height on SC phase and input magnetogram datasets, and quantitatively assesses the corresponding improvements in solar wind speed prediction at L1.
To properly interpret our results, it is necessary to first establish the relative strengths of the three SCs. Figure~\ref{fig:ssn} shows the monthly sunspot number during SC23, SC24, and SC25. From Figure~\ref{fig:ssn} we note the following:

\begin{itemize}
\item The light grey forward slashed region in Figure~\ref{fig:ssn} corresponds to the time period during which the low- and mid-latitude coronal hole area decreased prominently in SC23. During this interval, we reported a significantly reduced performance of the framework with conventional SS height \citep{kumar:2022}.
\item During the SC24 maximum, there exists a time period of exceptionally low solar activity  (marked by a backward slashed region) as shown in Figure~\ref{fig:ssn}.
\item It is also noted that the overall activity during the SC25 maximum is greater compared to the SC24 maximum.
\end{itemize}
For consistency and clarity, the same shading scheme as illustrated in Figure~\ref{fig:ssn} for respective periods is employed throughout the manuscript, particularly in Figures~\ref{fig:SC2324_main_ZPC_STD}, \ref{fig:SC25_main_ZPC_STD}, \ref{fig:SC24_main_ZPC_HMI}, and \ref{fig:SC25_main_ZPC_HMI}. 
\subsection{ Comparison Between the Framework Performance Using GONG ZPC and STD Maps}
In this sub-section, we compare the optimised SS height and model performance obtained using GONG ZPC and STD maps across SCs 23–25.
\subsubsection{SC 23-24} 
\begin{figure}[htbp]
    \centering
    \includegraphics[width=\linewidth]{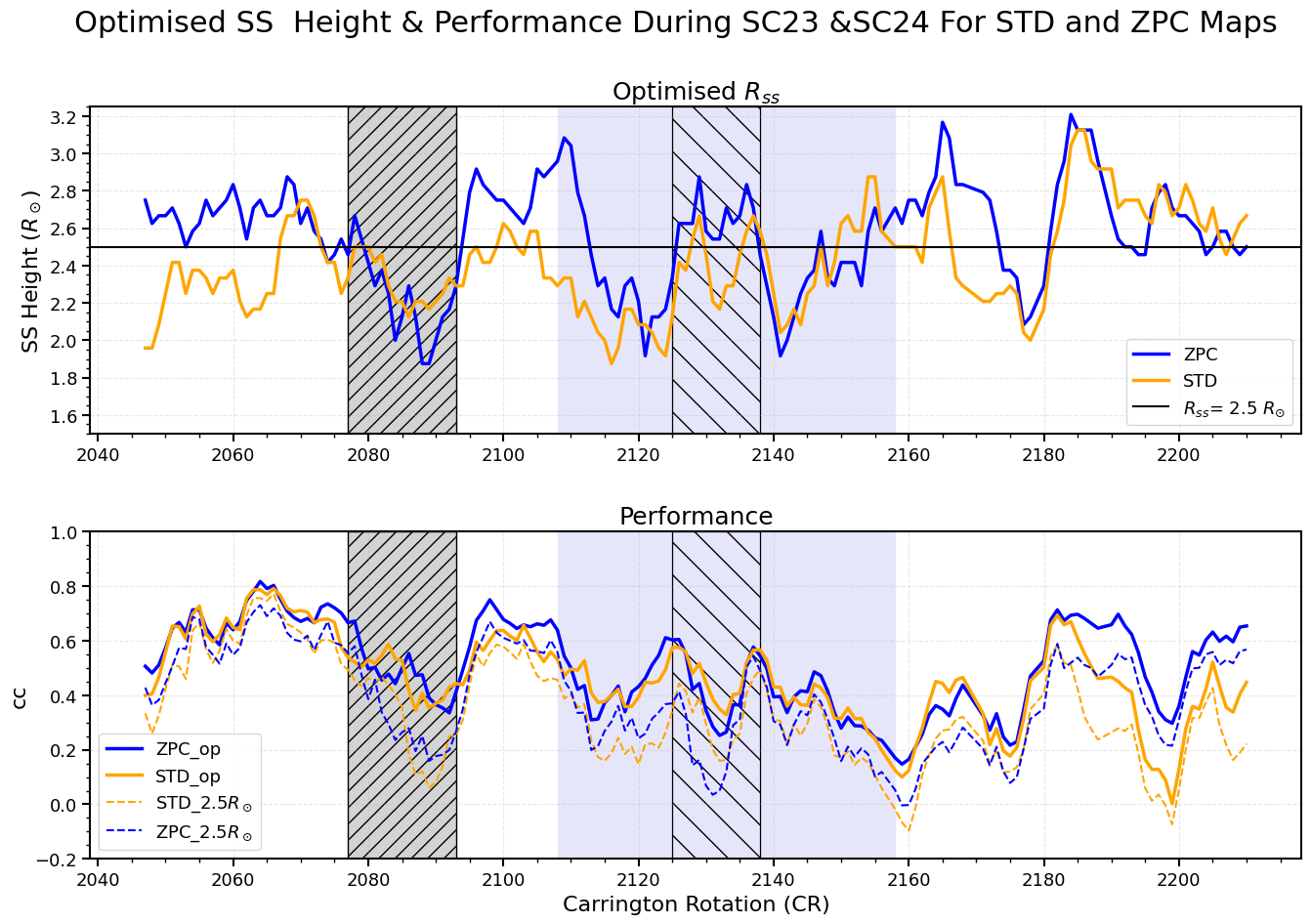}
    \caption{Optimized SS height and the corresponding performance of the framework for ZPC and STD maps during SC23 and SC24. Top panel shows 6 CR running average of the optimized SS height. Bottom panel shows the 6 CR running average of the performance of the framework.
 The solid blue and orange lines correspond to the performance of the optimised SS height for ZPC and STD maps, respectively, and the dashed lines show the corresponding performance for conventional SS height.}
    \label{fig:SC2324_main_ZPC_STD}
\end{figure}

\begin{figure}[htbp]
    \centering
    \includegraphics[width=\linewidth]{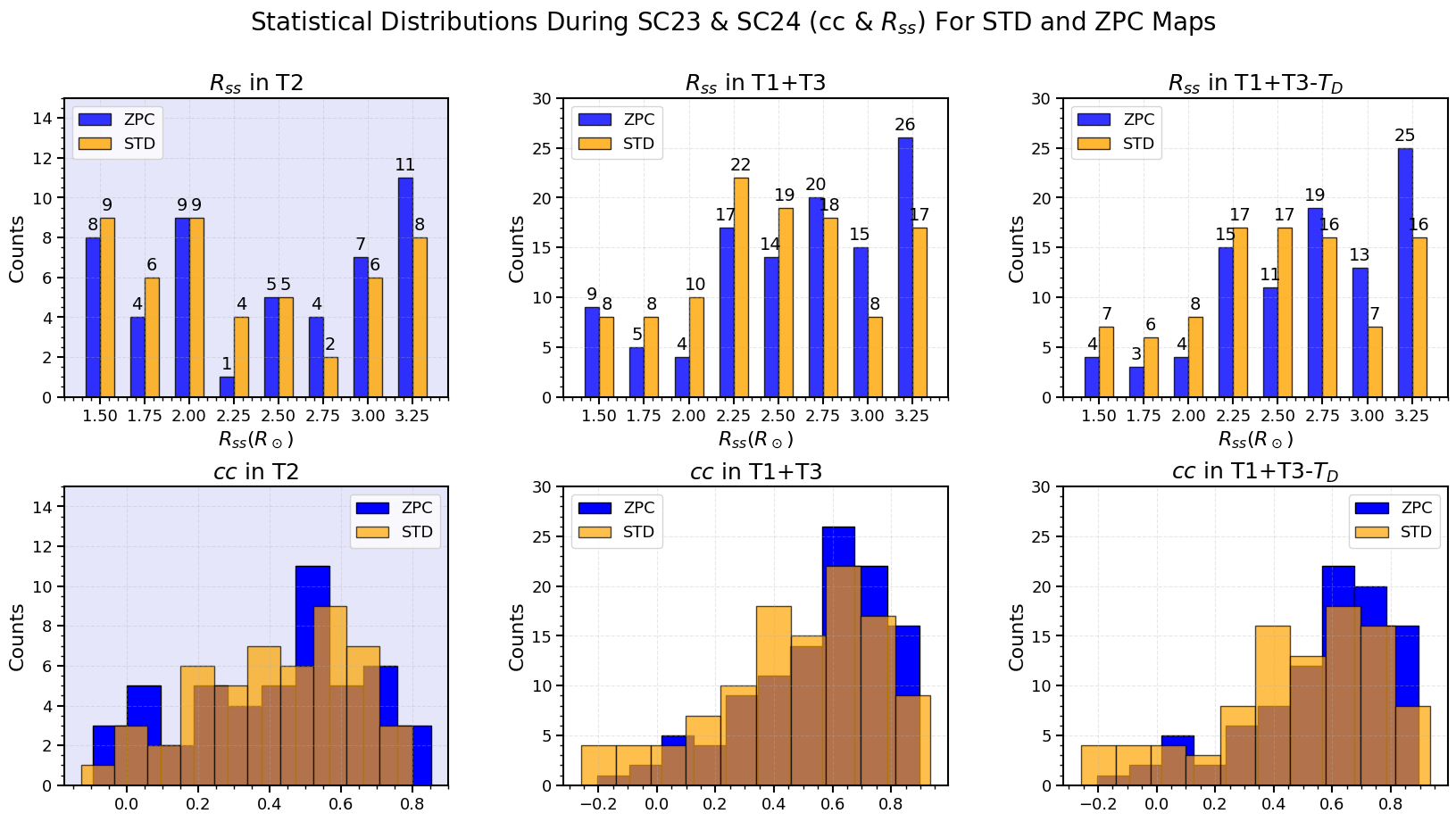}
    \caption{Distribution of the optimized SS height during different time periods of the SC23 $\&$  SC24 for STD and ZPC maps. Top panel shows the distribution of the optimized SS height. Bottom panel shows the distribution of performance of the framework corresponding to the optimized SS height. For the details of T1, T2 , T3 and $T_D$ refer to Table \ref{tab:time}.}
    \label{fig:SC2324_ST_ZPC_STD}
\end{figure}

Figure~\ref{fig:SC2324_main_ZPC_STD} shows the results of SS height optimisation for SC23 and SC24 using STD and ZPC synoptic maps.  Top panel of Figure~\ref{fig:SC2324_main_ZPC_STD} shows the running average of the optimised SS height in the PFSS model. Blue and orange curves in this plot show the value of the SS height for ZPC and STD maps, respectively, which provided the highest $cc$ for each CR when the modeled solar wind velocity was compared with the observed solar wind velocity at L1. As one can notice, for the minimum, deep minimum, descending, and ascending phases of SCs 23 and 24 (T1+T3, shown in the white region Figure~\ref{fig:ssn}), the optimised SS height is mostly greater than the conventional SS height (2.5 $R_\odot$), as indicated by the horizontal black line. Moreover, the distribution of the optimised SS heights (top right panel of Figure~\ref{fig:SC2324_ST_ZPC_STD}) during the minimum of SC23 and SC24 is clearly skewed towards the higher values of SS height ($\ge$ 2.5 $R_\odot$), with 3.25 $R_\odot$ as the best SS height for a larger number of CRs.   \\

For STD maps (solid orange curve in the top panel of Figure~\ref{fig:SC2324_main_ZPC_STD}), the trend of the curve remains the same, similar to results obtained using ZPC maps, except for the descending phase of SC23.  It is also noted that during the deep minimum phase of SC23 (light grey forward slashed region), a decrease in framework performance due to observational limitations (in measuring polar magnetic fields) was reported in \cite{kumar:2020} using STD GONG maps with conventional SS height. This decrease in framework performance is also evident in the bottom panel of Figure~\ref{fig:SC2324_main_ZPC_STD}, where we used different SS heights and ZPC maps. Therefore, this suggests that the observed decrease in performance is, in fact, due to observational limitations, as reported in our earlier study, rather than the choice of WSA parameters or the selection of the SS height. When the results from this interval ($T_D$) are excluded (Figure~\ref{fig:SC2324_ST_ZPC_STD}, top middle and right panels), the distribution of the optimized SS height becomes skewed towards higher values ($\ge 2.5 R_\odot$) of SS heights as compared to the lower values ($< 2.5 R_\odot$), e.g., 4 $\%$ and 3 $\%$ extra skewness for STD and ZPC maps respectively (Table \ref{tab:final}). However, this shift is very small and may not be statistically significant. 

Bottom panel of Figure~\ref{fig:SC2324_main_ZPC_STD} also shows the performance of the framework using the optimized and the conventional SS heights (2.5 $R_\odot$). It is noted that the difference in performance between the optimized and conventional SS heights is relatively small during the descending phase of SC23. In contrast, during SC24, the use of the optimized SS height leads to a comparatively significant improvement in the framework performance, as evident from the deviation of the solid blue and orange curves from dashed blue and orange curves. Moreover, towards the end of SC24, the framework performance obtained using ZPC maps is better than that obtained using STD maps. This behavior is also reflected in the distributions of the framework performance shown in the bottom panels of Figure~\ref{fig:SC2324_ST_ZPC_STD}.

During the SC24 maximum (lavender shaded regions in Figure~\ref{fig:SC2324_main_ZPC_STD}), no clear skewness is observed in the distribution of the optimized SS heights shown in top left panel of Figure~\ref{fig:SC2324_ST_ZPC_STD}. However, during SC24 maximum, there exists a distinct time interval (cross-hatched region in the top panel of Figure~\ref{fig:SC2324_main_ZPC_STD} and in Figure~\ref{fig:ssn}) in which higher values of SS heights ($\ge$ 2.5 $R_\odot$) perform better than lower values of SS heights ($< 2.5 R_\odot$). Interestingly, this interval coincides with the period of reduced solar activity during SC24 maximum (as shown in the backward slashed region Figure~\ref{fig:ssn}). When the contribution from this interval is excluded, the distribution of the optimized SS height shifts toward lower values of SS heights in the sample (Table \ref{tab:final}). For example, the percentage of $R_{ss} < 2.5,R_\odot$ for the ZPC and STD maps increased from $45\%$ to $56\%$ and from $54\%$ to $57\%$, respectively. Therefore, the apparent ambiguity in the SS height distribution during this period arises from the reduced solar activity within the SC24 maximum, suggesting a clear correlation between solar activity and the optimised SS height.

\begin{figure}[htbp]
    \centering
    \includegraphics[width=\linewidth]{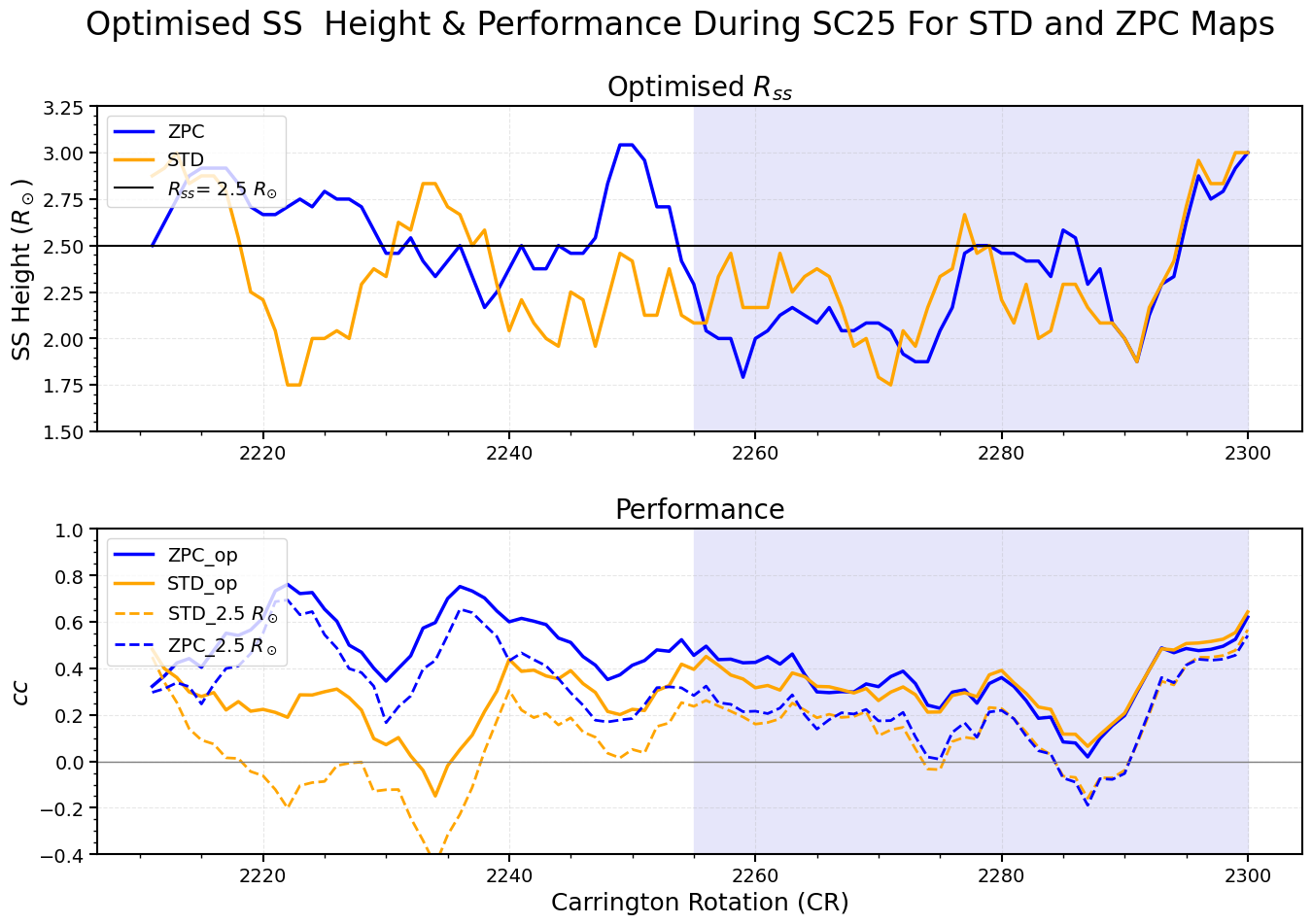}
    \caption{Similar to Figure~\ref{fig:SC2324_main_ZPC_STD}, but for SC25.}
    \label{fig:SC25_main_ZPC_STD}
\end{figure}

\begin{figure}[htbp]
    \centering
    \includegraphics[width=\linewidth]{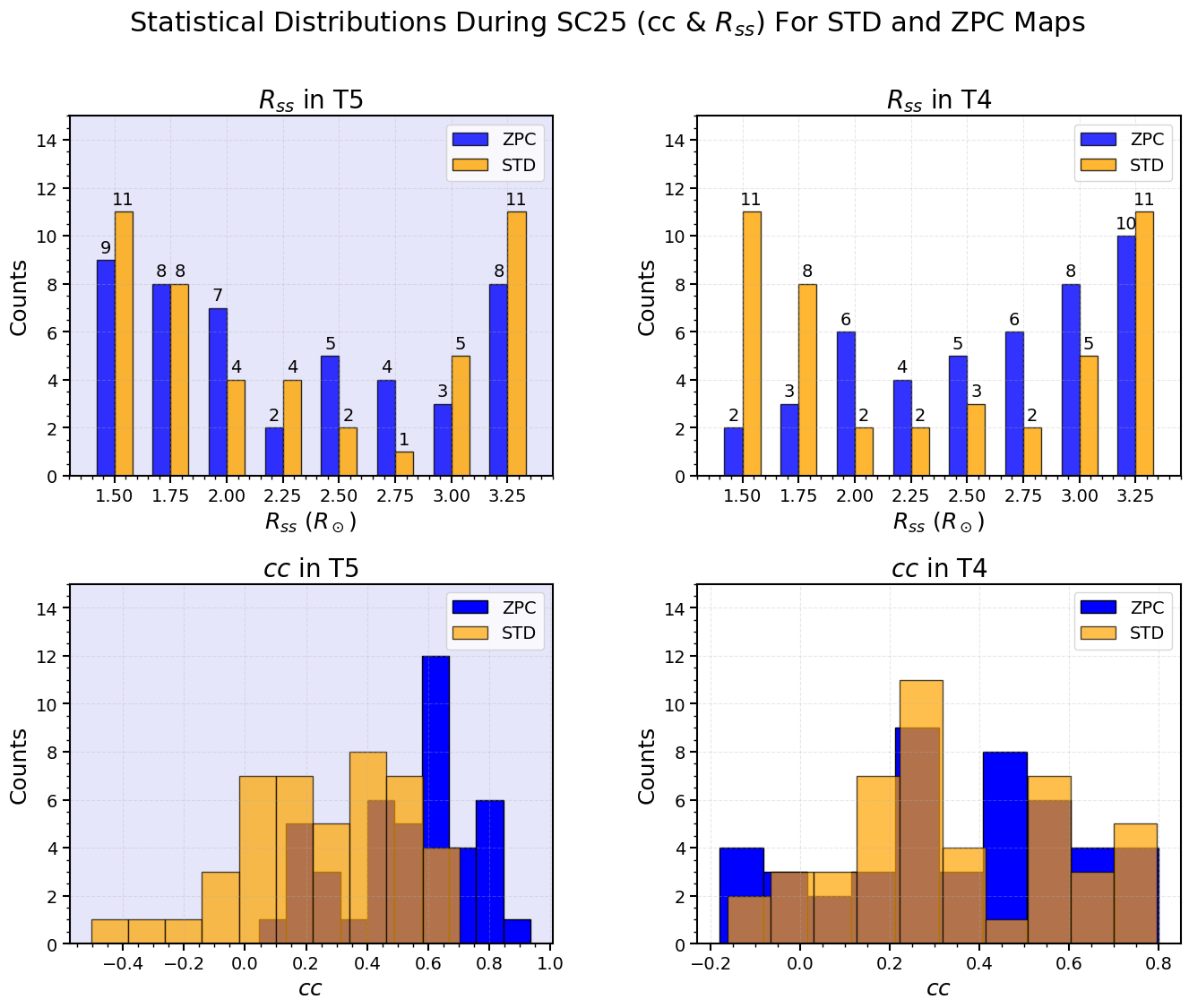}
    \caption{Similar to Figure~\ref{fig:SC2324_ST_ZPC_STD}, but for SC25. For the details of T4 and T5 refer to Table \ref{tab:time}. }
    \label{fig:SC25_ST_ZPC_STD}
\end{figure}

\begin{figure}[htbp]
    \centering
    \includegraphics[width=\linewidth]{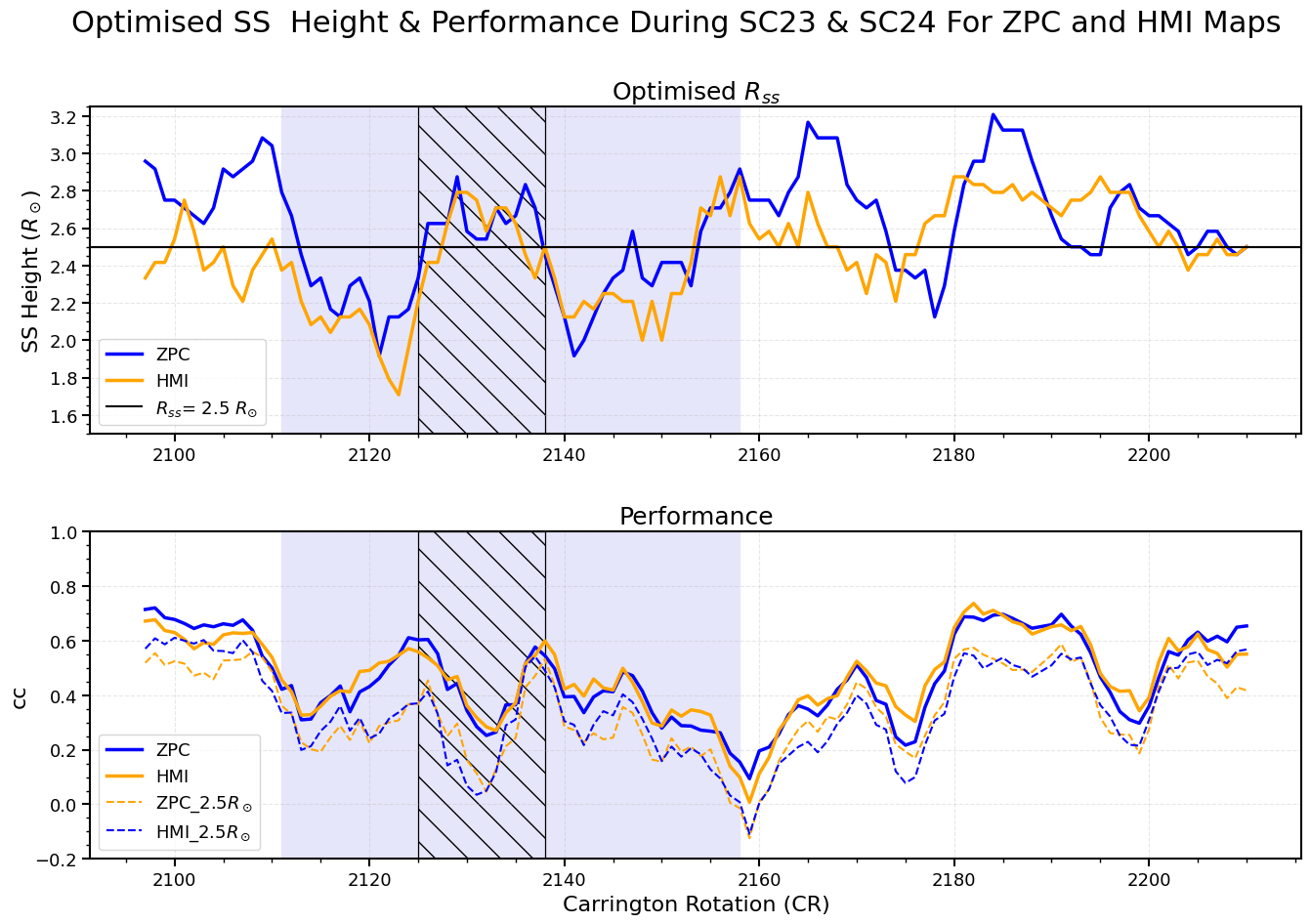}
    \caption{Optimized SS height and the corresponding performance of the framework for ZPC and HMI maps during SC23 $\&$ SC24. Top panel shows 6-CR running average of the optimized SS height. Bottom panle shows the 6-CR running average of the performance of the framework.
 The solid blue and orange lines correspond to the performance of the optimised SS height for ZPC and STD maps, respectively, and the dashed lines show the corresponding performance for conventional SS height. Color shading is exactly the same Figure~\ref{fig:ssn}}
    \label{fig:SC24_main_ZPC_HMI}
\end{figure}

\begin{figure}[htbp]
    \centering
    \includegraphics[width=\linewidth]{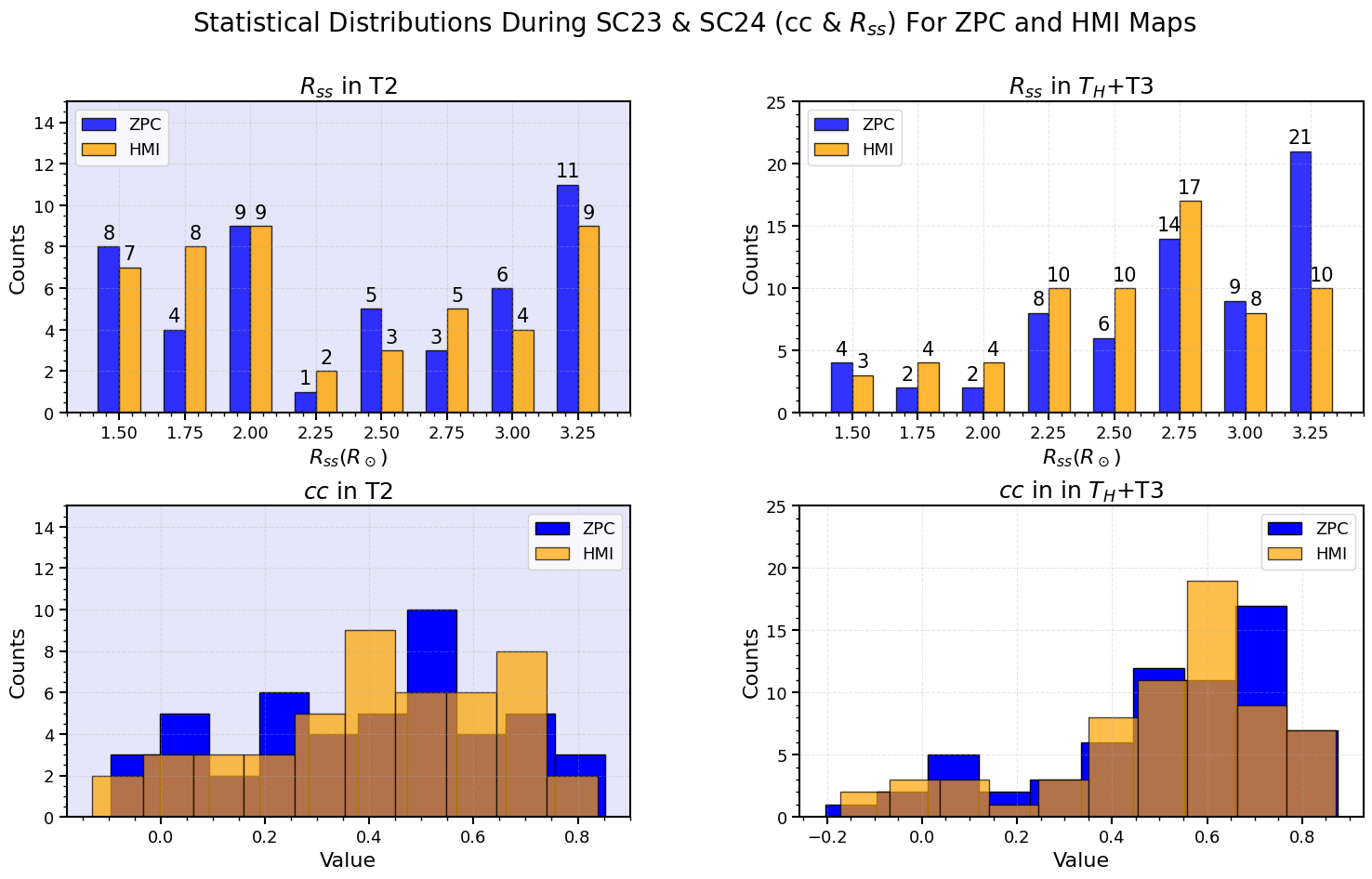}
    \caption{Distribution of the optimized SS height during different time periods of the SC23 $\&$  SC24 for STD and ZPC maps. Top panel shows the distribution of the optimized SS height. Bottom panel shows the distribution of performance of the framework corresponding to the optimized SS height. For the details of T2 , T3 and $T_H$ refer to Table \ref{tab:time}.}
    \label{fig:SC24_ST_ZPC_HMI}
\end{figure}

\begin{figure}[htbp]
    \centering
    \includegraphics[width=\linewidth]{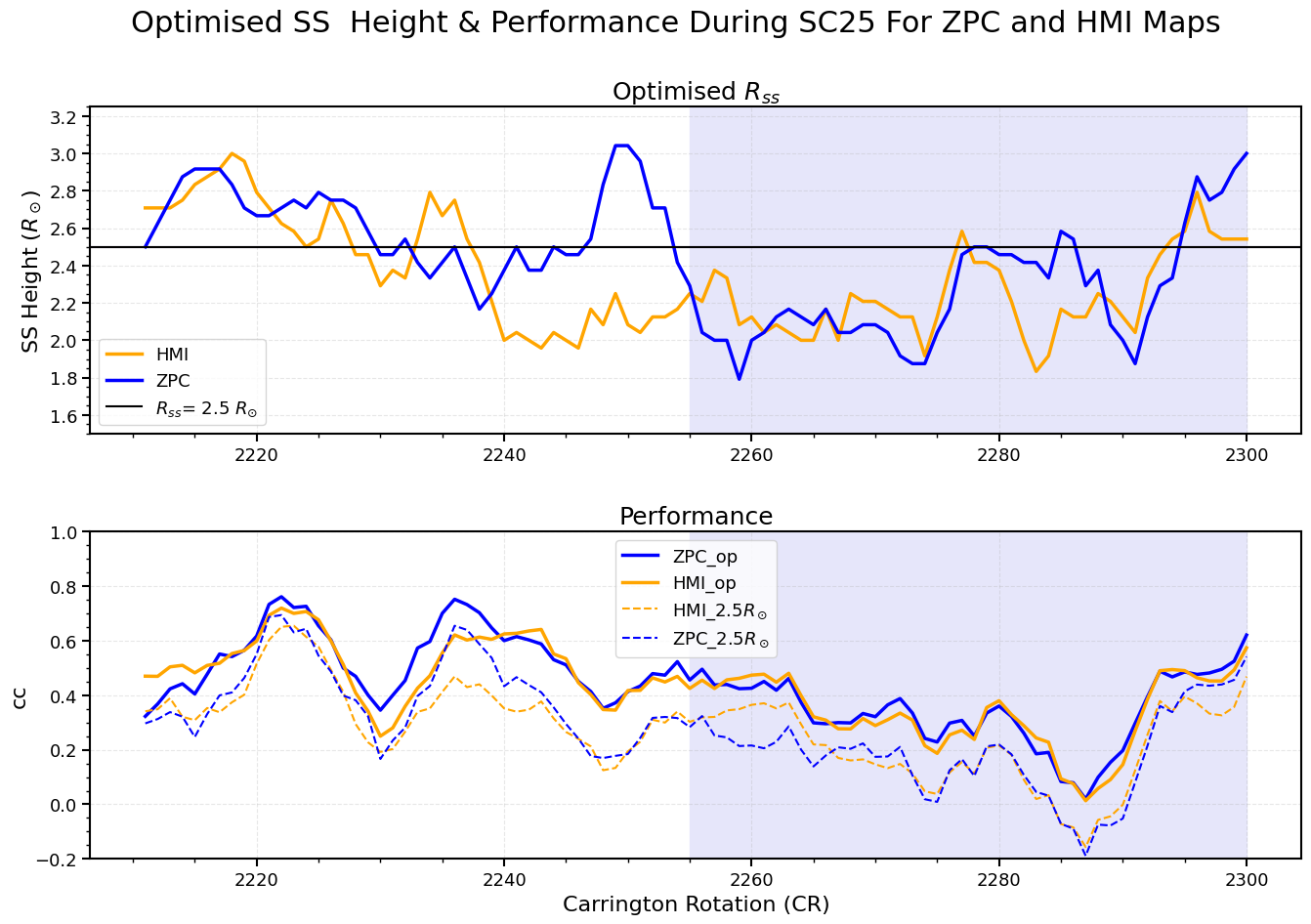}
    \caption{Optimized SS height and the corresponding performance of the framework for ZPC and HMI maps during SC25. Top panel shows 6-CR running average of the optimized SS height. Bottom panle shows the 6-CR running average of the performance of the framework. The solid blue and orange lines correspond to the performance of the optimised SS height for ZPC and STD maps, respectively, and the dashed lines show the corresponding performance for conventional SS height. Color shading is exactly the same Figure~\ref{fig:ssn}}
    \label{fig:SC25_main_ZPC_HMI}
\end{figure}

\begin{figure}[htbp]
    \centering
    \includegraphics[width=\linewidth]{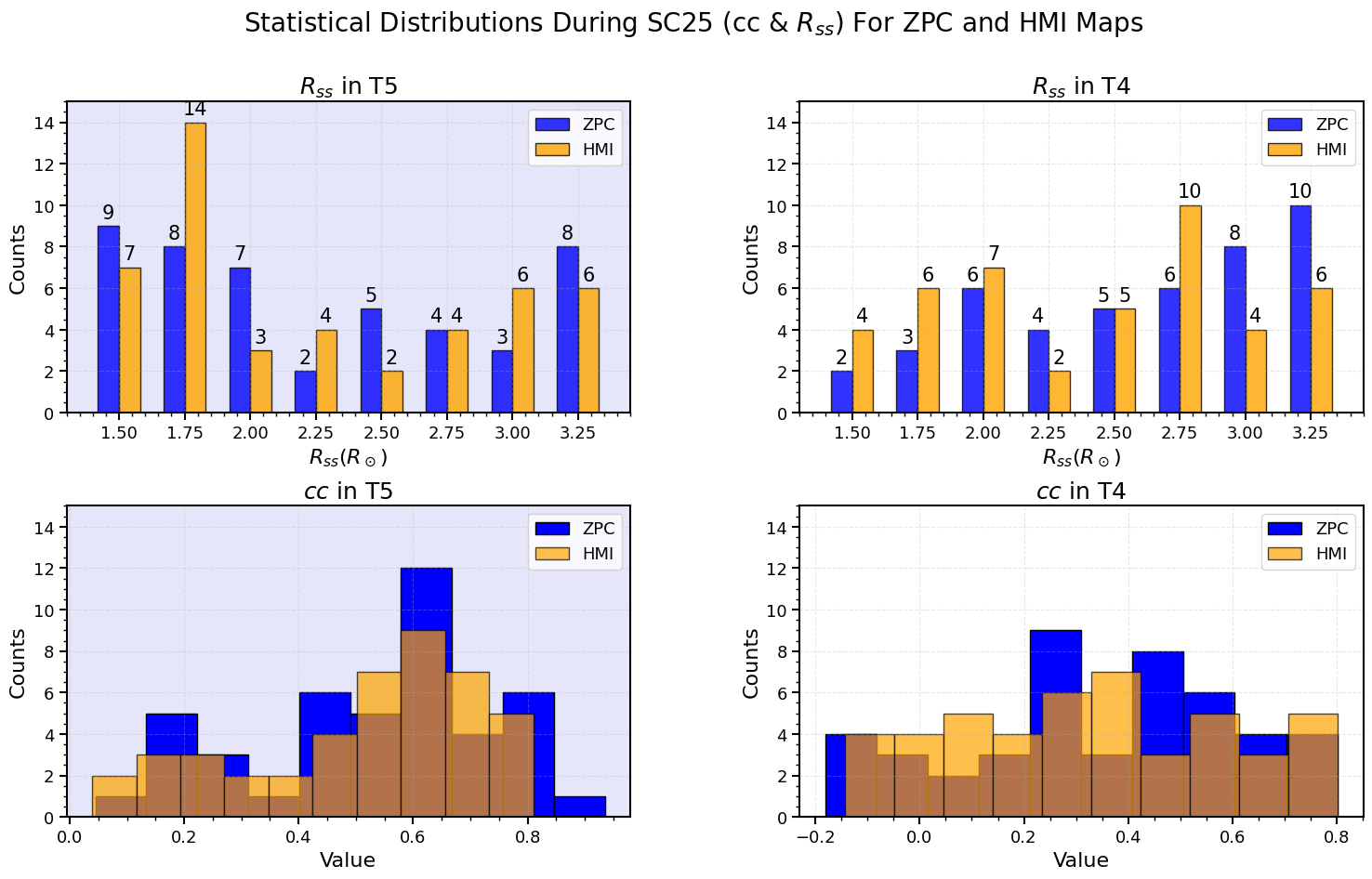}
    \caption{Distribution of the optimized SS height during different time periods of the SC25 for ZPC and HMI maps. Top panel shows the distribution of the optimized SS height. Bottom panel shows the distribution of performance of the framework corresponding to the optimized SS height. For the details of T4 and T5 refer to Table \ref{tab:time}.}
    \label{fig:SC25_ST_ZPC_HMI}
\end{figure}

\subsubsection{SC 25} Figure~\ref{fig:SC25_main_ZPC_STD} shows the results of optimised SS heights for SC25 using STD and ZPC synoptic maps. The very evident observation is that STD maps performed poorly as compared to the ZPC maps during SC25 ascending phase (bottom panel of Figure~\ref{fig:SC25_main_ZPC_STD}). The poor performance of the STD maps from 2006 to 2019 has been reported as the decreased quality of GONG maps, by \cite{nik:2019}, with their open flux calculation at L1. However, for solar wind velocity prediction of our case we found that the ZPC maps performed significantly better than the STD maps. 
We further found that beyond 2019 also ZPC maps out performed STD until the ascending phase of SC25, i.e., 2022. After this both ZPC and STD maps performed similarly in SC25 maximum (T5). 

Since ZPC maps perform significantly better than STD maps during most of the SC24 and SC25, we focus the discussion of the overall results based on ZPC maps only. During the SC25 maximum, the distribution of the optimized SS height shows a clear skewness toward lower SS heights, as illustrated in the top left panel of Figure~\ref{fig:SC25_ST_ZPC_STD} for the ZPC maps. This indicates that during SC25 maximum, on average, lower values of SS heights performed better than higher values of SS height. In contrast, during the ascending and minimum phases of SC25 (T4), the distribution of the optimized SS height is skewed toward higher values of SS heights as illustrated in the top right panel of Figure~\ref{fig:SC25_ST_ZPC_STD} for the ZPC maps.\\
It is also noteworthy that for SC25, the improvement in performance achieved by using the optimised SS height relative to the conventional SS height quantified by the separation between the solid and dashed curves in the bottom panel of Figure~\ref{fig:SC25_main_ZPC_STD}, is more pronounced than that observed during SC24. Furthermore, bottom panels of Figure~\ref{fig:SC25_ST_ZPC_STD} further demonstrate the superior performance of ZPC maps compared to STD maps.  

\subsection{Comparison Between the Framework Performance Using ZPC and HMI Maps}

In this section, we compare the optimised SS height and framework performance derived from ZPC and HMI synoptic magnetic field maps using the same PFSS–WSA modelling and validation methodology. We compare the results obtained using ZPC and HMI synoptic maps over their common observation period (CR2097 onward), i.e., SC24 and SC25 as shown in Figure~\ref{fig:SC24_main_ZPC_HMI}, Figure~\ref{fig:SC24_ST_ZPC_HMI}, Figure~\ref{fig:SC25_main_ZPC_HMI}, and Figure~\ref{fig:SC25_ST_ZPC_HMI}.

\subsubsection{SC 24} During SC24, the overall performance of the framework using ZPC and HMI synoptic maps is comparable. Moreover, the distributions of the optimized SS height are also largely similar (top panels of Figure~\ref{fig:SC24_main_ZPC_HMI} and Figure~\ref{fig:SC24_ST_ZPC_HMI}). One notable difference is that, during the SC24 maximum (T2), the distribution of the optimized SS height for HMI maps is skewed (55 $\%$) toward lower SS heights in contrast to the ZPC maps, which is skewed toward higher values of SS height (top left panel of Figure~\ref{fig:SC24_ST_ZPC_HMI}). During the SC24 ascending, declining and minimum phases, the ZPC and HMI  maps strongly suggest that higher SS height values yield better performance ($T_H$ +T3) top right panel of Figure~\ref{fig:SC24_ST_ZPC_HMI}. Nevertheless, both synoptic magnetic maps consistently indicate a preference for higher SS heights than the lower values of SS height in the sample during minimum and lower SS height  than 2.5 $R_\odot$ during maximum compared to the conventional SS height. Specifically, the ZPC maps suggest an optimal SS height of approximately $3.25$ $R_\odot$ during minimum of SC24, whereas the HMI maps favor a SS height of $2.75\,R_\odot$, as shown in top right panel of Figure~\ref{fig:SC24_ST_ZPC_HMI}.
\subsubsection{SC 25}  During SC25, results using HMI and ZPC maps both suggest similar optimised SS heights (top panel of Figure~\ref{fig:SC25_main_ZPC_HMI}). Although the difference is that during the ascending phase of the SC25 ($\approx$ CR2240-2255), optimised SS height for HMI maps is lower than the conventional SS height, whereas the ZPC suggested a higher SS height as compared to the lower values of SS heights in the sample. 
As observed during the SC24 maximum, during the SC25 maximum, the distribution of the optimised SS height for HMI maps is skewed toward lower SS heights than that for the ZPC maps, suggesting that the lower SS heights yield better performance for HMI data also during this phase (top left panel of Figure \ref{fig:SC25_ST_ZPC_HMI}).
During SC25, maximum  HMI maps suggested 1.75 $R_\odot$ as the best working SS height for a larger number of CRs (top left panel of Figure \ref{fig:SC25_ST_ZPC_HMI}), whereas it is 1.5  $R_\odot$ for ZPC maps.
Similar to the SC24 the ZPC maps suggest an optimal SS height of approximately $3.25$ $R_\odot$,  for larger number of CRs during minimum of SC24, whereas the HMI maps favor $2.75\,R_\odot$ (top right panel of Figure~\ref{fig:SC25_ST_ZPC_HMI}).

Further, it may be noted that the skewness of $R_{ss}$ distribution during the SC25 maximum (T5) toward lower values for the ZPC and HMI maps is more clear than during the SC24 maximum (T2). We attribute this to the higher activity level of SC25 maximum relative to SC24 maximum  (Figure~\ref{fig:ssn}).

We here note that HMI synoptic maps are rarely used for SS height optimisation in the context of their integration in solar wind velocity forecasting at L1. In this context, our study provides a first ever long-term comparison of GONG ZPC maps, which is widely used in WSA model, with the HMI maps.\\
Our study suggests a degradation of the GONG STD maps, as evidenced by decreased performance of the framework using STD maps as compared to the ZPC maps. This degradation of STD maps quality has also been reported by \cite{nik:2019} and \cite{li:2021}. Furthermore, our results are consistent with \cite{li:2021}, who found that for open magnetic flux calculation, HMI and GONG ZPC maps performed similarly, both of which performed better than GONG STD maps.

\subsection{Discussion}
Table~\ref{tab:final} summarises the results of the present study. It clearly shows that the optimised SS height ($R_{ss}$)  depends on both the SC phase and the type of the input magnetogram used. During  SC24 maximum, the distribution of $R_{ss}$ values is relatively balanced, with comparable fractions below and above $2.5\,R_{\odot}$. For SC24 maximum, removing the lower activity period (cross-hatched lavender region in the Figure~\ref{fig:ssn}) makes the results more skewed toward the lower values of SS height, e.g., 8$\%$ increase in the skewness for cases below $2.5\,R_{\odot}$ for HMI maps (Tabel \ref{tab:final}).
During ascending and minimum phases (SC23–24 and SC25), a large percentage of cases, e.g., 69 $\%$ for ZPC maps during T1+T3, exhibit $R_{ss} \geq 2.5\,R_{\odot}$, indicating a tendency toward higher optimal SS heights during low solar activity periods. It may be noted that removing the time period of decreased performance during SC23 minimum as reported in \cite{kumar:2022} further shifts the distribution slightly towards the higher values of SS height, e.g., $4\%$ and $3\%$ increase in the skewness for STD and ZPC maps respectively in T1+T3-$T_D$ as compared to T1+T3 as shown in table Table~\ref{tab:final}.

At this point, we would like to emphasize that, in this study, we analyze the optimized SS height results in the view of actual indicators of solar activity, i.e., sunspot numbers, rather than relying on selected CRs to represent different phases of the SC. This approach helps explain some previously reported deviations from the general trend. 
For example, the relatively low-activity interval within the SC24 maximum closely corresponds to the CR selection (e.g., CR2137) used by \cite{huang:2024}. While they reported a higher SS height than the conventional value, consistent with our findings, we identify this period as a local minimum rather than a true SC24 maximum (see Figure~\ref{fig:ssn}). 
Furthermore, some of their selected CRs (e.g., CR2106 and CR2167) lie close to the boundaries of the SC maximum phase, introducing ambiguity in phase classification. In this context, our statistical approach, based on a longer time scale, provides a more robust interpretation. 

Additionally, differences in results based on the input synoptic magnetic field maps are also evident: ZPC dataset consistently shows a stronger preference for higher $R_{ss}$ values during ascending and minimum phases, compared to HMI and STD datasets. This suggests that the inferred optimal SS height is sensitive to the choice of input magnetogram and coronal model assumptions.

It is important to acknowledge the significant spread of the optimised SS height distribution in our results that goes against the general trend, particularly during the solar maximum period. Here, it is important to discuss the definition of SS height and the applicability of the PFSS model. In the original formulation of the PFSS model, the SS is defined as the height above which the thermal energy density of the coronal plasma exceeds the transverse magnetic field energy density \citep{sch:1969}. The relative radial gradients of these two quantities determine the effective height of the SS. Under this physical interpretation, we expect that SS is not perfectly spherical, owing to the inherently non-uniform and asymmetric nature of the solar atmospheric environment  \citep{sch:1969}. However, in the standard community implementation of the PFSS model, a global spherical source surface is assumed. Therefore, the results of SS height optimisation must be interpreted only in an average, global, and long-term sense. Moreover, the PFSS model is only expected to work during lower activity periods, because the current free approximation of the solar corona holds only during the minimum period. Therefore, the non-global nature of the "true source surface" and the limited time period of applicability of the PFSS model to the solar cycle might actually contribute to the spread of the optimised SS height distribution across different phases.

It is important to note that during solar maximum, the coronal magnetic field is predominantly multipolar and the corona is hotter, whereas during solar minimum, the field is largely bipolar and the corona is comparatively cooler. Owing to the multipolar nature of the magnetic field at solar maximum, the transverse magnetic field energy density is expected to decrease more rapidly with height than during solar minimum, when the large-scale field is more dipolar. Consequently, based on the physical definition of the source surface, the effective SS height is expected to be lower globally during solar maximum and higher during solar minimum when averaged over long timescales. The arguments mentioned above can justify the overall trend observed in our study and those of previous studies that agree with our conclusions.


\begin{table*}[ht]
\centering
\caption{Percentage distribution of optimised $R_{\rm ss}$ for different SCs, datasets, and solar activity phases. The results are grouped into two categories: $R_{\rm ss} < 2.5~R_{\odot}$ and $R_{\rm ss} \geq 2.5~R_{\odot}$. For the details of the regions mentioned in this table, see Figure~\ref{fig:ssn}.}
\begin{tabular}{l l l c c}
\hline
Solar Cycle & Phase & Dataset & $R_{\rm ss} < 2.5~R_{\odot}$ (\%) & $R_{\rm ss} \geq 2.5~R_{\odot}$ (\%) \\
\hline
SC24 & T2 & ZPC & 45 & 55 \\
SC24 & T2- $T_M$ & ZPC & 56 & 44\\
SC24 & T2 & HMI & 57 & 43 \\
SC24 & T2- $T_M$ & HMI & 65 & 35 \\
SC24 & T2 & STD & 57 & 43 \\
SC24 & T2- $T_M$ & STD & 54 & 46 \\
SC23-24 & T1+T3 & ZPC & 31 & 69 \\
SC23-24 & T1+T3 & STD & 44 & 56 \\
SC23-24 & T1+T3-$T_D$ & ZPC & 28 & 72 \\
SC23-24 & T1+T3-$T_D$ & STD & 40 & 60 \\
\hline
SC25 & T5 & HMI & 61 & 39 \\
SC25 & T4 & HMI & 43 & 57\\

SC25 & T5 & ZPC & 56 & 44 \\
SC25 & T4  & ZPC & 34 & 66\\
\hline
\end{tabular}
\label{tab:final}
\end{table*}

\section{Conclusion}

In this work, we present a comprehensive analysis of the source surface (SS) height, the only free parameter in the widely used PFSS model, particularly in the context of its application to solar wind prediction at L1. We analysed all the available synoptic magnetogram data sets spanning nearly three solar cycles (SC23, SC24, and SC25), from both space-based and ground-based observatories, i.e., GONG and SDO/HMI. We further explore the impact of using two different GONG magnetogram products (ZPC and STD) on the optimisation of the SS height. Our study not only represents one of the longest-term investigations of SS height variability in the PFSS model, but it also uses one of the most accurate magnetogram observations (GONG and HMI) currently being used in space weather forecasting frameworks. We specifically focused on SS optimisation for solar wind speed prediction at L1. 

We found that employing higher values of the SS height during periods of low solar activity leads to improved solar wind velocity predictions at L1, whereas lower SS heights are more suitable during periods of high solar activity. However, the optimized SS heights during SC maxima exhibit significantly greater scatter compared to those during minima, where a clearer distinction in optimal values is observed. This increased variability can be attributed partly to the limited applicability of the PFSS model during high-activity periods, and partly to the presence of localized low-activity intervals within the solar cycle 24 maximum. We also reported that optimised SS height might also depend upon the choice of the magnetogram, however, the relative trend remains similar for different phases of SC.

We also reported that GONG ZPC and HMI synoptic maps performed similarly, and both outperformed GONG STD maps for solar wind velocity prediction at L1. The absolute values of the optimised SS height vary with the SC phase and differ from one SC to another, due to the uniqueness of every SC.

Despite the differences, our results are broadly consistent with those reported in earlier studies \citep{lee:2011,Arden:2014,nik:2019,Bena:2024,kumar:2025}, particularly in terms of the relative variation in the optimal SS height across different phases of the SC. These earlier investigations primarily relied on estimates of the open magnetic flux measured at L1. Moreover, previous efforts to optimise the SS height were limited in scope, as they either employed a single type of magnetic field map over a specific time interval or relied on a narrow set of observational constraints such as open flux at L1 or coronal hole area to validate model outputs \citep{lee:2011,Arden:2014,nik:2019}. In contrast, our study examines SS height optimisation within the WSA framework using an extended dataset, including the full archive of available GONG synoptic magnetic field maps covering 254 CRs and HMI synoptic magnetic field maps covering 204 CRs.

Our study differs from earlier studies based on modeled open flux from PFSS at L1, which show a direct correlation between open flux and $R_{ss}$ \citep{lee:2011,Arden:2014,nik:2019}. However, in our case, it is not straightforward to correlate the solar wind velocity at L1 and SS height.  It is due to the fact that the WSA velocity depends on the field-line properties, such as the expansion factor ($f_s$) and distance from the coronal hole boundary ($\theta_b$), rather than on the total open flux, which increases as $R_{ss}$ decreases. Additional complexity may come from the use of more advanced heliospheric propagation methods (e.g., HUX). Nevertheless, our results remain broadly consistent with previous studies, as both are ultimately governed by the extrapolated coronal magnetic structures.

Our study, along with previous studies \cite{Arden:2014,nik:2019,zhang:2023,kumar:2025}, suggests different SS heights for different phases of SC in more broader sense only, i.e., overall distribution shifts or in the running averages.  However, some exceptions to this trend have been reported in certain studies \citep{bad:2020,huang:2024,shoda:2025}, which may be attributed to the different methodology or different proxies used to compare with the observations from the PFSS model.\\

In summary, our long-term study improves our understanding of how the height of the source surface influences solar wind modelling across different phases of the SC. This knowledge can contribute to refining predictions of solar wind velocity and its implications for Earth, i.e., using a slightly higher SS height than 2.5 $R_\odot$ near the minimum.  Despite the overall trends obtained from the optimized spherical source-surface height, we further propose adopting a potential-field model with a non-symmetric, non-spherical source surface to better account for the intrinsically non-spherical nature of the solar atmosphere environment.

\begin{acknowledgments}
 The authors have used pfsspy \citep{pfss:2020}. The data used in this study are all publicly available. We want to acknowledge the GONG group for magnetic synoptic maps available on \url{https://gong2.nso.edu/archive/patch.pl?menutype=zeroPoint#step2}. 
HMI synoptic magnetic field maps can be freely obtained from \url{http://jsoc.stanford.edu/HMI/Magnetograms.html}.
  Solar wind velocity data is available on https://omniweb.gsfc.nasa.gov/form/dx1.html. The sunspot data is courtesy of WDC-SILSO, Royal Observatory of Belgium, Brussels (\url{https://www.sidc.be/SILSO/datafiles}). We acknowledge the Sunpy community \citep{sunpy_community2020}.  
\end{acknowledgments}

\bibliography{manuscript_clean}{}
\bibliographystyle{aasjournal}

\end{document}